\documentstyle[12pt,epsf,rotating,axodraw]{article}

\catcode`@=11
\def\citer{\@ifnextchar
[{\@tempswatrue\@citexr}{\@tempswafalse\@citexr[]}}

\def\@citexr[#1]#2{\if@filesw\immediate\write\@auxout{\string\citation{#2}}\fi
  \def\@citea{}\@cite{\@for\@citeb:=#2\do
    {\@citea\def\@citea{--\penalty\@m}\@ifundefined
       {b@\@citeb}{{\bf ?}\@warning
       {Citation `\@citeb' on page \thepage \space undefined}}%
\hbox{\csname b@\@citeb\endcsname}}}{#1}}
\catcode`@=12

\newcommand{\MS}{\overline{\rm MS}}
\newcommand{\non}{\nonumber}
\newcommand{\beq}{\begin{equation}}
\newcommand{\eeq}{\end{equation}}
\newcommand{\beqn}{\begin{eqnarray}}
\newcommand{\eeqn}{\end{eqnarray}}

\begin{document}

\renewcommand{\thefootnote}{\fnsymbol{footnote}}

\begin{flushright}
DESY 99--060 \\
hep-ph/9905469 \\
May 1999 \\[1.0cm]
\end{flushright}

\begin{center}
{\Large \bf $W$ Boson Production at NLO\footnote{Contribution to the
Workshop '{\it Monte Carlo Generators for HERA Physics}', DESY, 1998/99}
} \\[0.5cm]

{\sc Michael Spira\footnote{Heisenberg-Fellow}} \\[0.5cm]

{\it II.\ Institut f\"ur Theoretische Physik\footnote{Supported in part
by the EU FF Programme under contract FMRX-CT98-0194 (DG 12 - MIHT)},
Universit\"at Hamburg, D-22761 Hamburg, Germany} \\[0.5cm]
\end{center}
\begin{abstract}
\noindent
We discuss $W$ boson production at HERA including NLO QCD corrections. A
detailed comparison with previous work is presented.
\end{abstract}

\renewcommand{\thefootnote}{\arabic{footnote}}
\setcounter{footnote}{0}

\section{Introduction}
The c.m.\ energy $\sqrt{s}\approx 300$ GeV of the $ep$ collider HERA 
\cite{hera}
is sufficiently large to produce on-shell $W$ bosons. Since the
production cross sections for the processes $e^\pm p \to e^\pm W + X$ reach
values of about 1 $pb$ at HERA, the number of $W$ events allows to
study the production mechanisms of $W$ bosons in some detail and to probe
the existence of anomalous $WW\gamma$ trilinear
couplings \citer{heratri,dubinin}. Moreover, $W$ boson production
represents an important SM
background to several new physics searches such as the measurement
of isolated high energy muons
\cite{muon}. In order to observe possible discrepancies between the
observations and Standard Model (SM) predictions, the latter have
to be sufficiently accurate and reliable.
This is not guaranteed for the available leading order
calculations \cite{baur,dubinin,lo}
of $W$ boson production. Clearly, for an unambiguous test of
anomalous contributions, it is necessary to extend the previous
analyses to NLO accuracy. A first step in this direction will be presented
in this contribution.

\section{QCD Corrections}
\subsection{Leading Order}
The production of $W$ bosons at $ep$ colliders is mediated by photon, $Z$
and $W$ exchange between the electron/positron and the hadronic part of
the process. It is useful to distinguish two
regions, the DIS regime at large $Q^2$ and the photoproduction regime
at small $Q^2$, $Q^2$ being the square of the transferred momentum.
The photoproduction cross section can be calculated by
convoluting the Weizs\"acker-Williams photon spectrum,
\beq
f_{\gamma/e}(x) = \frac{\alpha}{2\pi} \left\{ \frac{1+(1-x)^2}{x} \log
\frac{Q^2_{max}}{Q^2_{min}} - 2 m_e^2 x \left( \frac{1}{Q^2_{min}} -
\frac{1}{Q^2_{max}} \right) \right\},
\label{eq:wwa}
\eeq
with the cross section for $\gamma q \to q' W$:
\beq
\sigma (ep \to W+X) = \int_{M_W^2/s}^1 d\tau \sum_q \frac{d{\cal
L}^{\gamma q}}{d\tau}
\hat \sigma(\gamma q \to q' W; \hat s = \tau s)
\eeq
where
\beq
\frac{d{\cal L}^{\gamma q}}{d\tau} = 
\int_\tau^1 \frac{dx}{x} f_{\gamma/e}(x)\ q_p\left(\frac{\tau}{x},\mu_F^2\right)
\eeq
is the photon-quark luminosity,
$\alpha$ denotes the QED coupling, $m_e$ the electron mass, and
$Q^2_{min}, Q^2_{max}$ the minimal and maximal values of the photon
virtuality $Q^2$. The function
$q_p(x,\mu_F^2)$ is the quark density of the proton at
the momentum fraction $x$ and the factorization scale $\mu_F$.
In order to separate photoproduction from
the DIS region we impose an
angular cut of $\theta_{cut} = 5^o$ on the outgoing lepton, 
which corresponds
to an energy-dependent cut
\beq
Q^2_{max} = \frac{E_e^2 (1-x)^2 \theta_{cut}^2 + m_e^2 x^2}{1-x},
\eeq
$E_e$ being the initial lepton energy \cite{q2max}.
The minimal value of $Q^2$ is fixed by kinematics,
\beq
Q^2_{min} = m_e^2 \frac{x^2}{1-x},
\eeq
where negligible higher order terms in the electron mass $m_e$ 
have been omitted.

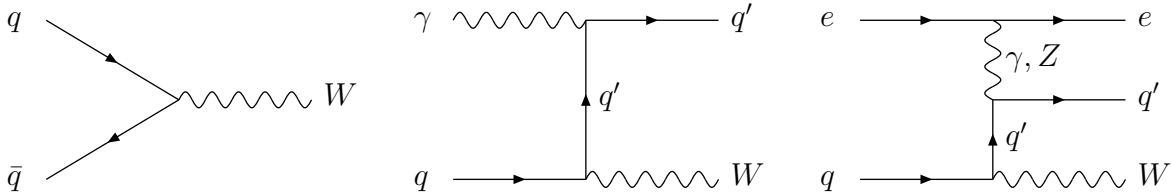
\begin{figure}[hbtp]
\begin{center}
\begin{picture}(120,80)(20,10)

\ArrowLine(0,80)(50,50)
\ArrowLine(50,50)(0,20)
\Photon(50,50)(100,50){3}{5}

\put(-15,78){$q$}
\put(-15,18){$\bar q$}
\put(105,48){$W$}

\end{picture}
\begin{picture}(120,80)(-10,10)

\ArrowLine(0,20)(50,20)
\ArrowLine(50,20)(50,80)
\ArrowLine(50,80)(100,80)
\Photon(50,20)(100,20){3}{5}
\Photon(0,80)(50,80){3}{5}

\put(-15,78){$\gamma$}
\put(-15,18){$q$}
\put(55,48){$q'$}
\put(105,78){$q'$}
\put(105,18){$W$}

\end{picture}
\begin{picture}(120,80)(-40,10)

\ArrowLine(0,80)(50,80)
\ArrowLine(50,80)(100,80)
\ArrowLine(0,20)(50,20)
\ArrowLine(50,20)(50,50)
\ArrowLine(50,50)(100,50)
\Photon(50,20)(100,20){3}{5}
\Photon(50,50)(50,80){3}{3}

\put(-15,78){$e$}
\put(-15,18){$q$}
\put(55,33){$q'$}
\put(55,63){$\gamma,Z$}
\put(105,78){$e$}
\put(105,48){$q'$}
\put(105,18){$W$}

\end{picture}  \\
\caption[]{\label{fg:dia} \it Typical diagrams of $W$ boson production at
HERA: resolved, direct and DIS mechanism.}
\end{center}
\vspace*{-0.5cm}
\end{figure}

While the treatment of the DIS region is straightforward [a typical
contribution is shown in the third diagram of Fig.~\ref{fg:dia}], the small
$Q^2$ region requires to include the contribution of the hadronic
component of the photon giving rise to
$W$ production via the standard
Drell-Yan mechanism [first diagram of Fig.~\ref{fg:dia}]. 
In fact, this is the dominant production mechanism.

The leading direct photon process $\gamma q\to q'W$ 
[a typical contribution is depicted by the
second diagram of Fig.~\ref{fg:dia}] develops
a singularity when the final state quark $q'$ becomes collinear with
the initial state photon. This singularity has to be subtracted and absorbed
in the corresponding quark density of the photon. We have worked in the
$\MS$ scheme using dimensional
regularization. The subtraction of the collinear pole introduces the
factorization scale $\mu_F$ in  the photonic quark density. The renormalized
result for the direct contribution can be cast into the form \cite{NRS}
\beqn
\hat \sigma_{LO}^{dir} & = & \frac{G_F M_W^2 \alpha}{2\sqrt{2} \hat s}
\left\{ e_{q'}^2 \left[ -2 [z^2+(1-z)^2] \log\left(\frac{\mu_F^2 z}{M_W^2
(1-z)^2}\right) +1+6z-7z^2 \right] \right. \\
& & \hspace{1.8cm} + 2 e_{q'} e_W \left[3 (1-z^2) +
4(1+z^2)\log z \right] \non \\
& & \hspace{1.8cm} \left. + e_W^2 \left[ \frac{1-z}{z} (4+5z+7z^2) +
(8+4z+4z^2) \log z \right] \right\} \non
\eeqn
where $G_F$ denotes the Fermi constant, $M_W$ the $W$ mass and $e_{q'},
e_W$ the electric charges of the scattered quark $q'$ and $W$ boson,
i.e.\ $e_W = e_q-e_{q'} = \pm 1$. The variable $z$ is defined 
to be $z=M_W^2/\hat
s$. The subtracted direct component is accounted for by the 
resolved process $q'\bar q\to W$, where one initial quark comes from the
proton and the other from the photon in the collinear regime. The
corresponding production cross section is given by
\beq
\sigma^{res} (ep\to W+X) = \int_{M_W^2/s}^1 d\tau
\sum_{q,q'} \frac{d{\cal L}^{q'\bar q}}{d\tau}
\hat \sigma^{res}(q'\bar q \to W; \hat s = \tau s)
\eeq
with the quark-antiquark luminosity
\beq
\frac{d{\cal L}^{q'\bar q}}{d\tau} =
\int_\tau^1 \frac{dx}{x} \int_x^1 \frac{dy}{y} f_{\gamma/e}(y)\
\left[q'_\gamma \left(\frac{x}{y}, \mu_F^2 \right)\
\bar q_p\left(\frac{\tau}{x}, \mu_F^2 \right)\
+ \bar q_\gamma \left(\frac{x}{y}, \mu_F^2 \right)\
q'_p\left(\frac{\tau}{x}, \mu_F^2 \right) \right]
\eeq
and the partonic cross section at leading order [$z=M_W^2/\hat s$]
\beq
\hat \sigma_{LO}^{res} (q' \bar q\to W) = \frac{\sqrt{2}G_F\pi}{3}
\delta(1-z).
\eeq
\begin{figure}[hbtp]
\vspace*{0.5cm}

\hspace*{2.0cm}
\begin{turn}{-90}%
\epsfxsize=8cm \epsfbox{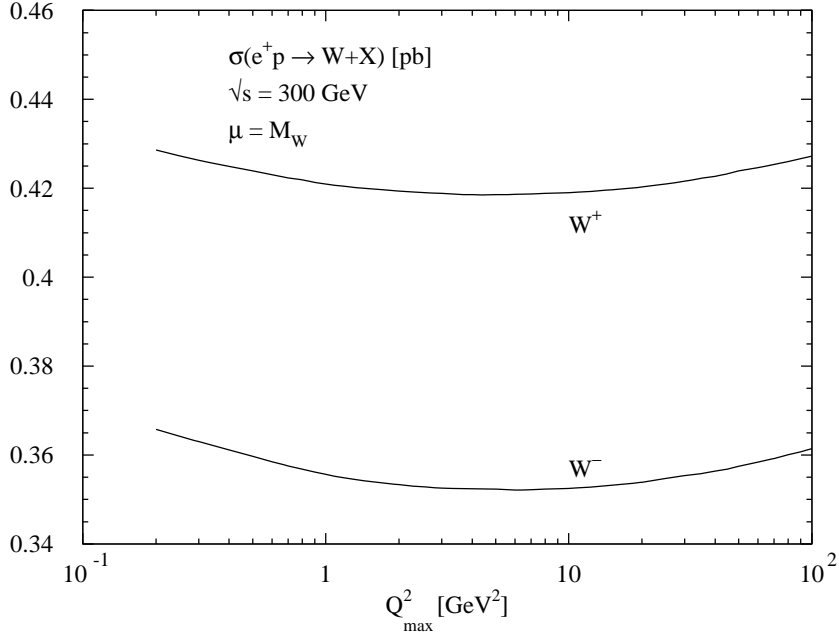}

\end{turn}
\vspace*{0.0cm}

\caption[]{\it \label{fg:q2} Dependence of the total leading order $W^\pm$
production cross sections on the cut $Q^2_{max}$ which separates the DIS and
photoproduction regimes, after adding the DIS, direct and resolved
contributions.}
\end{figure}

The DIS, direct and resolved contributions add up to the total $W$ 
production cross section.
The consistency of the calculation requires that the
dependence on the specific value of the cut $Q^2_{max}$, which separates
the DIS and photoproduction regimes, should be small. This dependence
is presented in Fig.~\ref{fg:q2} for
the LO $W^+$ and
$W^-$  cross sections. It can be seen that the residual dependence is
less than about 3\% and thus indeed sufficiently small.

\subsection{Next-to-leading Order}
For the dominant resolved part we have evaluated the QCD corrections in the
$\overline{\rm MS}$ scheme \cite{NRS}. Since the resolved process coincides
with the Drell-Yan production of $W$ bosons, the final renormalized result for
the total resolved cross section is given by \cite{drellyan}
\beqn
\sigma^{res}(ep\to W+X) & = & \sigma^{res}_{LO} + \Delta\sigma^{res}_{q\bar q} +
\Delta\sigma^{res}_{qg}
\nonumber \\
\Delta\sigma^{res}_{q\bar q} & = & \frac{\sqrt{2}G_F\pi}{3}\
\frac{\alpha_s(\mu^2_R)}{\pi} \int_{M_W^2/s}^1 d\tau \sum_{q,q'}
\frac{d{\cal L}^{q'\bar q}}{d\tau}~z~\omega_{q\bar q}(z) \nonumber \\
\Delta\sigma^{res}_{qg} & = & \frac{\sqrt{2}G_F\pi}{3}\
\frac{\alpha_s(\mu^2_R)}{\pi}\int_{M_W^2/s}^1
d\tau \sum_{q,\bar q} \frac{d{\cal L}^{qg}}{d\tau}~z~\omega_{qg}(z)
\eeqn
with the coefficient functions [$z=M_W^2/(\tau s)$]
\beqn
\omega_{q\bar q}(z) & = & -P_{qq}(z) \log \frac{\mu_F^2 z}{M_W^2}
\nonumber \\
& & + \frac{4}{3}\left\{ 2[\zeta_2-2]\delta(1-z)
+ 4\left(\frac{\log(1-z)}{1-z}\right)_+ - 2(1+z)\log(1-z)
\right\} \nonumber \\
\omega_{qg}(z) & = & -\frac{1}{2} P_{qg}(z) \log \left(
\frac{\mu_F^2 z}{(1-z)^2 M_W^2} \right) + \frac{1}{8}\left\{ 1+6z-7z^2
\right\} \, .
\eeqn
Here, $P_{ij}(z)$ denote the usual Altarelli-Parisi splitting functions
\cite{altpar}, and $\alpha_s(\mu_R^2)$ is the strong coupling at the
renormalization scale $\mu_R$. The quark-gluon luminosity is
given by
\beq
\frac{d{\cal L}^{qg}}{d\tau} =
\int_\tau^1 \frac{dx}{x} \int_x^1 \frac{dy}{y} f_{\gamma/e}(y)\
\left[q_\gamma \left(\frac{x}{y}, \mu_F^2 \right)\
g_p\left(\frac{\tau}{x}, \mu_F^2 \right)\
+ g_\gamma \left(\frac{x}{y}, \mu_F^2 \right)\
q_p\left(\frac{\tau}{x}, \mu_F^2 \right) \right]
\eeq
with $g_{\gamma,p}(x,\mu_F^2)$ denoting the gluon densities of the photon
and proton, respectively.
\begin{figure}[hbtp]
\vspace*{0.5cm}

\hspace*{2.0cm}
\begin{turn}{-90}%
\epsfxsize=8cm \epsfbox{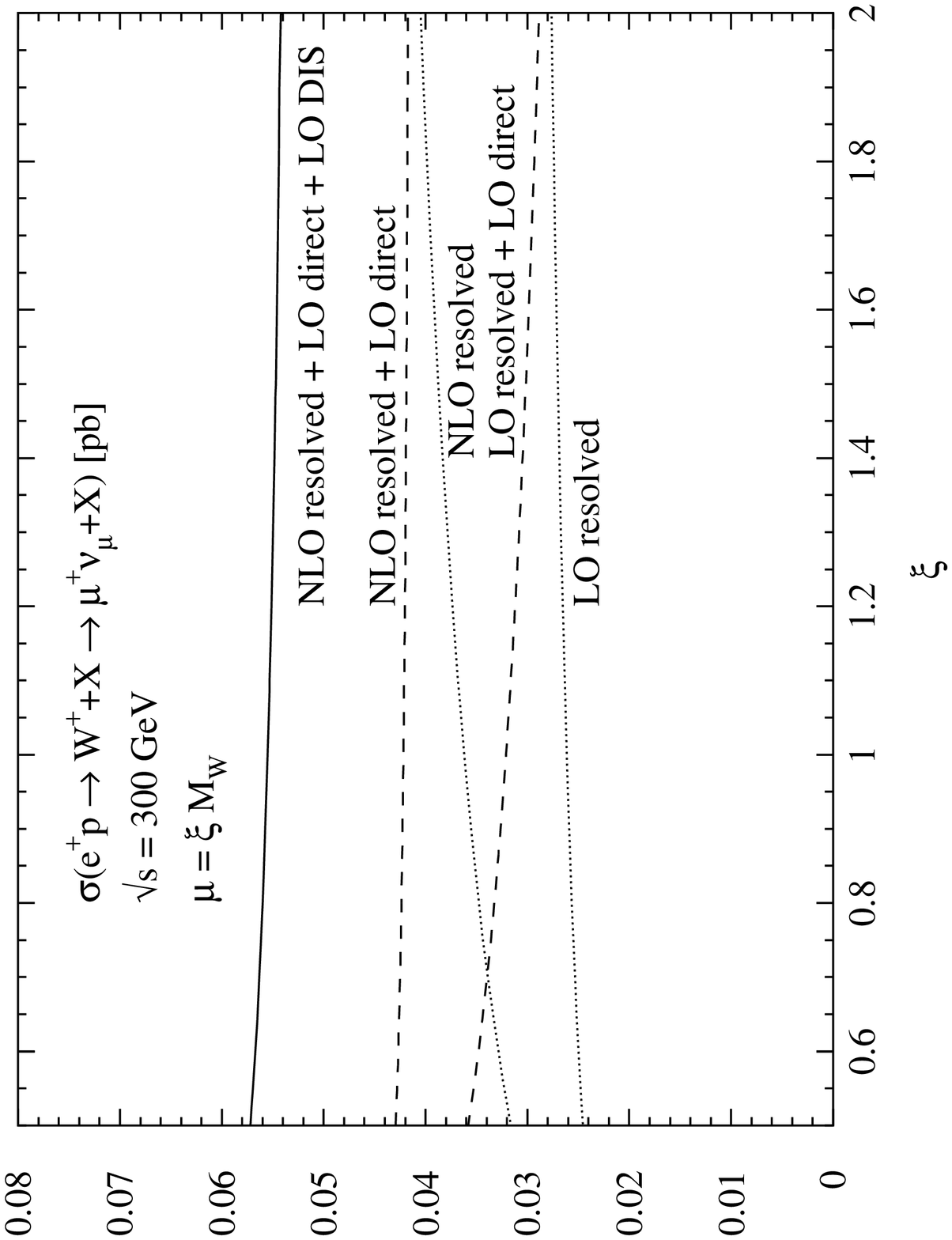}
\end{turn}
\vspace*{0.5cm}

\hspace*{2.0cm}
\begin{turn}{-90}%
\epsfxsize=8cm \epsfbox{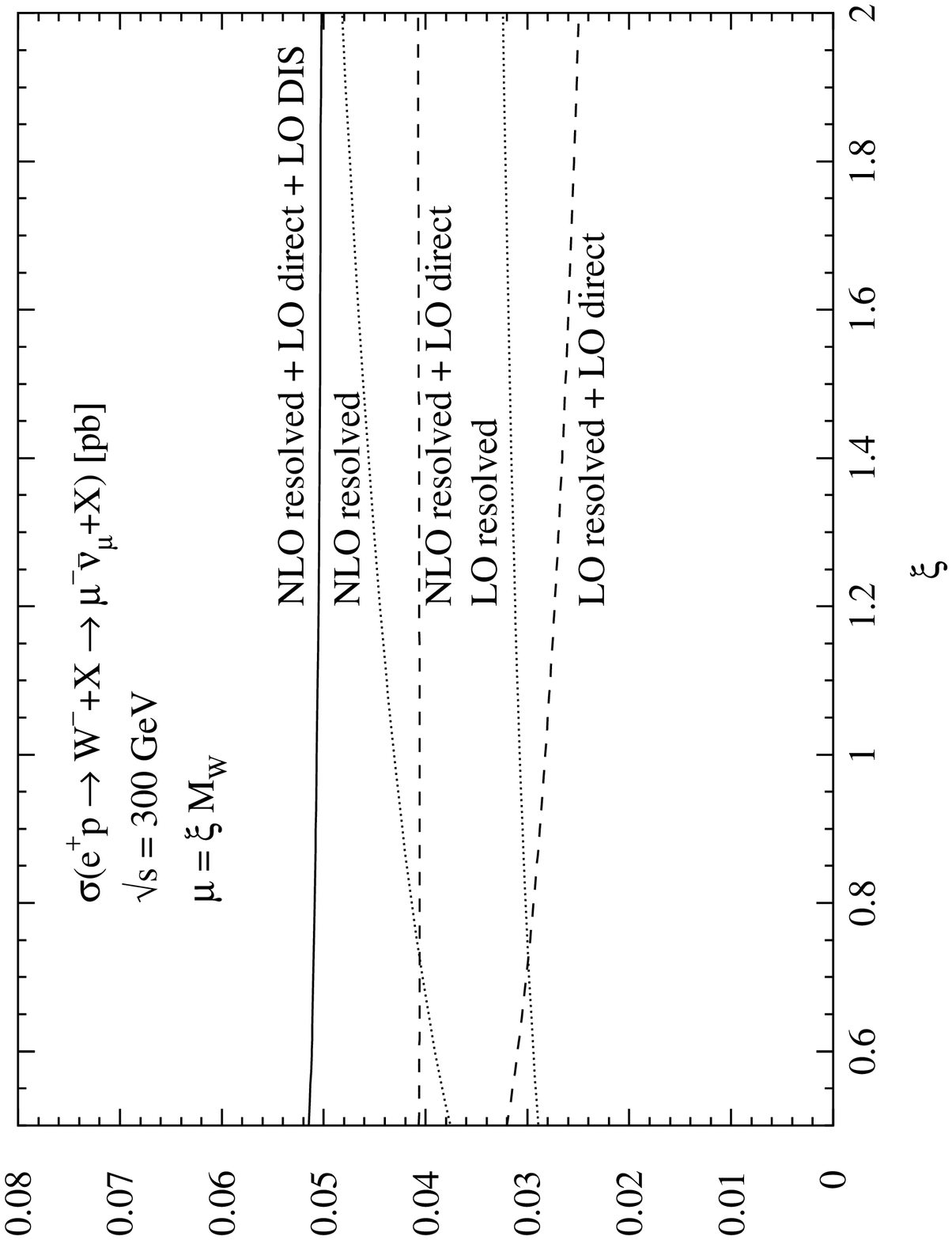}
\end{turn}
\vspace*{-0.0cm}

\caption[]{\label{fg:scale} \it Dependence of the individual
contributions to $W^+$ (upper plot) and $W^-$ (lower plot) production
on the renormalization and 
factorization scale $\mu = \mu_F = \mu_R = \xi M_W$.
The full curves represent the final predictions for the
total cross section of $W^\pm$ production in $e^+p$ collisions. We have
chosen CTEQ4M \cite{cteq} and ACFGP \cite{acfgp} parton densities for
the proton and the photon,
respectively. The strong coupling constant is taken at NLO with
$\Lambda_5 = 202$ MeV. An angular cut of $5^o$ is introduced for the
separation of photoproduction and deep inelastic scattering (DIS).}
\end{figure}
Taking the cross section at the values
$\mu_R = \mu_F = M_W$ for the renormalization and factorization scales,
the QCD corrections enhance the resolved contribution by about 40\% for
$W^+$ and $W^-$ production. In order to demonstrate the theoretical
uncertainties, the renormalization/factorization scale dependence of the
individual contributions to the processes $e^+ p \to W^\pm + X \to \mu^\pm
\!\!\stackrel{\mbox{\tiny (}{\scriptscriptstyle -}\mbox{\tiny
)}}{\nu}_{\!\!\mu} + X$
are presented in Fig.~\ref{fg:scale} for HERA conditions
[including the branching ratio $BR(W^\pm\to \mu^\pm \!\! \stackrel{\mbox{\tiny
(}{\scriptscriptstyle -}\mbox{\tiny )}}{\nu}_{\!\!\mu}) = 10.84\%$].
One can clearly see that the scale
dependence in the sum of direct and resolved contributions is
significantly reduced, once the NLO corrections to the resolved part are
included. The full curves show  the total sum of NLO resolved, LO direct
and LO DIS contribution, that is our prediction of 
the total $W^\pm$ production cross
sections. The residual scale dependence
is about 5\%. Since the remaining dependence on $Q^2_{max}$ is of similar
size, the total theoretical uncertainty is estimated to be less than about
10\%. 

\begin{figure}[hbtp]
\vspace*{0.5cm}

\hspace*{2.0cm}
\begin{turn}{-90}%
\epsfxsize=8cm \epsfbox{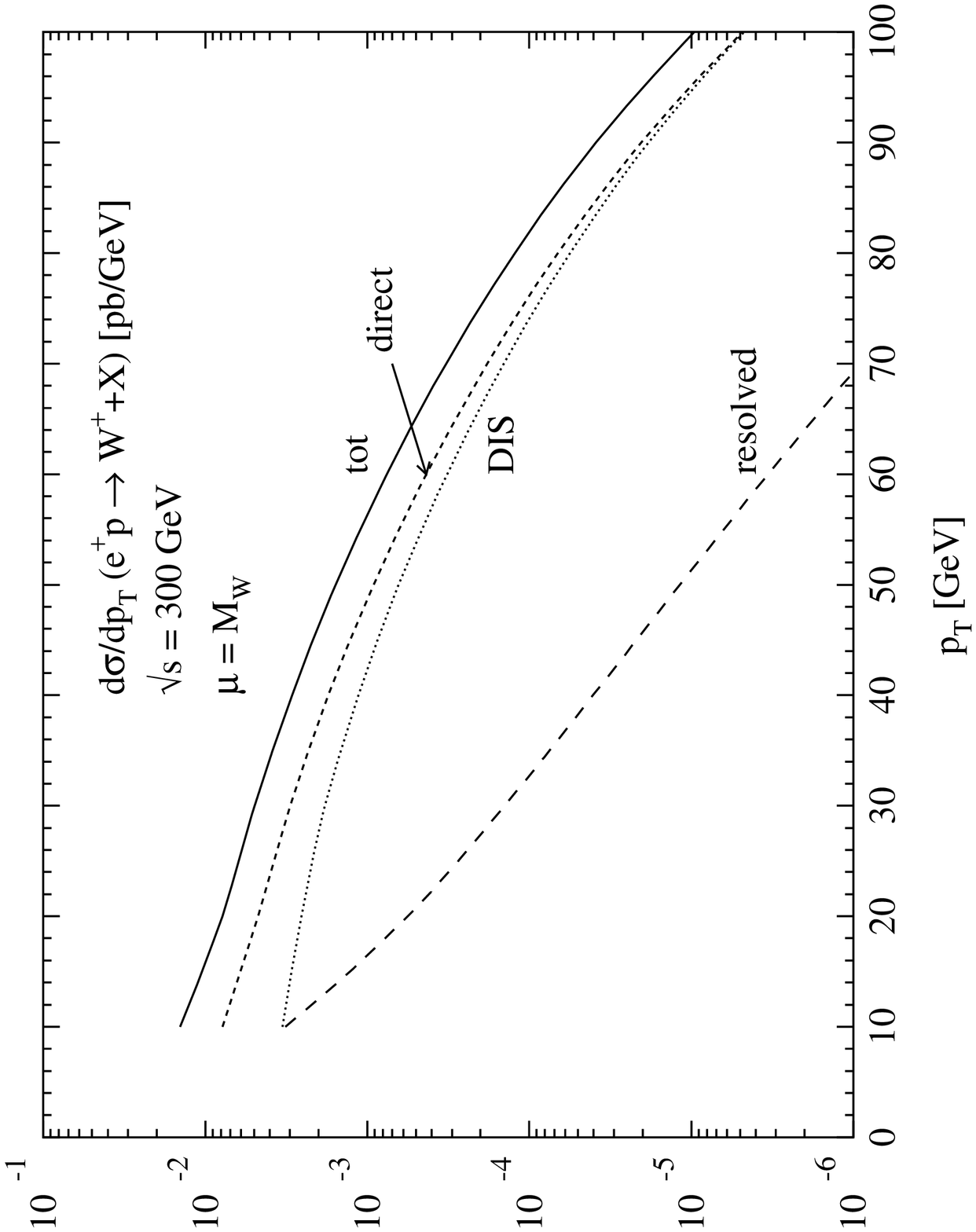}
\end{turn}
\vspace*{0.0cm}

\caption[]{\it \label{fg:pt} Transverse momentum distribution of $W^+$
bosons at HERA. The full curve shows the total $p_T$ distribution, while
the broken lines exhibit the individual DIS, direct
and resolved contributions.}
\end{figure}

The radiation of an additional gluon also generates a finite transverse
momentum $p_T$ of the $W$ bosons produced via the resolved Drell-Yan process. 
At sufficiently low $p_T$ this may be expected to modify 
the total $p_T$ distribution.
As can be inferred from Fig~\ref{fg:pt}, at $p_T$ values
below 20 GeV the resolved contribution amounts to about 5\% and more
of the total $p_T$ distribution of the $W$ bosons, while at larger
values of $p_T$ it falls off steeply\footnote{The fraction of resolved
$W$ events in all events with $p_T > 15$ GeV is about 5\%.}.
There, the $p_T$ spectrum is dominated by the direct photon mechanism and
DIS which becomes more and more important as $p_T$ increases.
Moreover, for $p_T$ values below about 15--20 GeV multi soft gluon radiation
should become important. This would require resummation 
in order to obtain a finite result. The
description of $W$ production in this small $p_T$ regime is 
beyond the scope of the
present analysis.

\section{Comparison with Earlier Results}
Our approach differs from the analysis of Ref.~\cite{baur} in several
aspects:
\begin{description}
\item[(i)~~] Whereas in Ref.~\cite{baur} the DIS and 
photoproduction regimes are
separated by a cut on the $u$-channel momentum transfer in the $\gamma^*
q$ subprocess [see second diagram of Fig.~\ref{fg:dia}], here
these two regions are separated by a more conventional cut
in the photon virtuality $Q^2$.

\item[(ii)~] In Ref.~\cite{baur} photoproduction is treated 
in the DIS$_\gamma$ scheme, making use of the
quark densities extracted from the structure function $F_2^\gamma$
as measured in $\gamma^* \gamma \to q\bar q$. In contrast,
our analysis is
carried out in the conventional $\MS$ scheme.

\item[(iii)] In Ref.~\cite{baur} an
approximation of the Weizs\"acker-Williams spectrum of quasi-real
photons is used which only includes the first logarithmic term of the curly
bracket in eq.(\ref{eq:wwa}). Moreover, the input for  
$Q^2_{max/min}$ differs from our choice.

\item[(iv)~] In Ref.~\cite{baur}, the full amplitude for off-shell $W$ 
production is computed including the
leading amplitudes for non-resonant 4-fermion final states. We 
have only considered 
on-shell $W$ production.
\end{description}

If we make similar approximations for the Weizs\"acker-Williams
spectrum and work in the DIS$_\gamma$ scheme,
we are able to reproduce the results of Ref.~\cite{baur} 
within less than 10\%. The
residual differences can be attributed to the different treatment of the
DIS and photoproduction regimes and to non-resonant contributions.

\section{Conclusions}
We have presented predictions for $W$ boson production at HERA including
the QCD corrections to the dominant resolved photon mechanism. Working in
the conventional $\MS$ scheme we find that the QCD corrections
enhance the resolved contributions by about 40\% at the nominal
renormalization/factorization scale $\mu_R=\mu_F=M_W$, and thus
have a sizeable effect on the total $W$ production rate.
In addition, the NLO corrections reduce the residual
scale dependence of the total cross section to a level of about 5\%. 
Taking into account also the
variation with the cut separating the DIS and photoproduction
regimes, the total theoretical uncertainty is estimated to be
smaller than about 10\%. This does not yet include the uncertainties 
from the parton densities of the photon and proton. 
In spite of the dominance of the resolved photon mechanism in the total
cross sections, gluon radiation in the resolved photon process
hardly affects the $W$ transverse momentum spectrum at $p_T$
values above about 15 GeV.

Our approach differs significantly from earlier analyses
particularly in the treatment of the separation between the DIS and
photoproduction regimes. Nevertheless, the resulting total $W$
production cross sections differ by less than about 10\%. \\

\noindent {\bf Acknowledgements.}\\
I would like to thank P.\ Nason and R.\ R\"uckl for their pleasant
collaboration and G.\ Altarelli, M.\ Dubinin, M.\ Kuze, D.\ Waters and
D.\ Zeppenfeld for useful discussions.

\end{document}